\documentclass[12pt]{article}
\pdfoutput=1
\usepackage{Max}

\allowdisplaybreaks[3]
\title{Recounting Special Lagrangian Cycles in Twistor Families of K3 Surfaces \\
\large{Or: How I Learned to Stop Worrying and Count BPS States}}
\author[1]{Shamit Kachru}
\author[2]{Arnav Tripathy}
\author[1]{Max Zimet}
\affil[1]{Stanford Institute for Theoretical Physics,

Stanford University, Stanford, CA 94305 USA

~}
\affil[2]{Department of Mathematics,

Harvard University, Cambridge, MA 02138 USA}
\date{}

\begin{document}

\maketitle

\begin{abstract}
We consider asymptotics of certain BPS state counts in M-theory compactified on a K3 surface. Our investigation is parallel to (and was inspired by) recent work in the mathematics literature by Filip \cite{filip:counting}, who studied the asymptotic count of special Lagrangian fibrations of a marked K3 surface, with fibers of volume at most $V_*$, in a generic twistor family of K3 surfaces. We provide an alternate proof of Filip's results by adapting tools that Douglas and collaborators have used \cite{douglas:statistics,douglas:countingFlux,douglas:critical,douglas:distributions,douglas:nonSUSY,douglas:algGeom} to count flux vacua and attractor black holes.  We similarly relate BPS state counts in 4d ${\cal N}=2$ supersymmetric gauge theories to certain counting problems in billiard dynamics and provide a simple proof of an old result in this field.
\end{abstract}

\newpage
\tableofcontents
\hypersetup{linkcolor=blue}

\section{Introduction}

Determining the existence and properties of BPS states has a long and storied history.  Key applications include those to 
uncovering duality symmetries of string theory \cite{Sen:1994fa} and understanding the dynamics of supersymmetric quantum field theories
\cite{sw:theory1,sw}.  
Asymptotics of counts of BPS states at large mass and charge are particularly relevant to understanding 
microstate counts underlying the Bekenstein-Hawking entropy of simple
black holes \cite{stromingerVafa}.

\bigskip
The usual approach in such counting problems is to fix some supercharges and count states that are BPS with respect to those supercharges. However, in theories with sufficient supersymmetry, there is another natural question which, to our knowledge, has been less investigated: counting states which are BPS with respect to {\it any} choice of supercharges at a fixed point in the moduli space. Indeed, this is the more natural question; as we will find below, this approach to counting detects states which the usual index computations miss. In a mathematical context, this question has been recently addressed in \cite{filip:counting, bergeron:slag}, where a formula determining the large-volume asymptotics of special Lagrangian $T^2$ fibrations of $K3$ (as a function of the fiber volume) was proposed.

\bigskip
Let us introduce and motivate this problem from the perspectives of both physics and mathematics. Consider M-theory compactified to 7 dimensions on K3. We may obtain BPS states by wrapping M2-branes on supersymmetric 2-cycles, which are simply holomorphic curves with respect to some complex structure \cite{becker:slag,vafa:k3inst}.\footnote{The existence of massive BPS states might seem strange in light of the fact that the vacuum preserves the smallest amount of supersymmetry allowed in seven dimensions. However, in many dimensions besides $d=4$, including $d=7$, minimally supersymmetric theories may contain massive BPS states \cite{strathdee:SUSY}.} Equivalently, thanks to the hyper-K\"ahler structure of K3, we may regard such a cycle as special Lagrangian with respect to some choice of K\"ahler form, $\omega$.
That is, if our curve is $C$, and the holomorphic 2-form is denoted by $\Omega$, we have
$$\omega\vert_C = 0,~~{\rm vol}(C) = {\rm Re}\left(e^{i\theta}\Omega\right)\vert_C,~~{\rm Im}\left(e^{i\theta}\Omega\right)\vert_C = 0$$
for some $\theta$.
 Therefore, we are interested in counting 2-cycles which are special Lagrangian (with any phase $\theta$) with respect to an arbitrary K\"ahler form. Actually, since these curves $C$ generically have moduli spaces (even for a fixed choice of K\"ahler form) this is still not a well-defined problem: to count BPS states of the M2-branes we should really study the cohomology of these moduli spaces, and in particular their Euler characteristics. If we restrict to genus one 2-cycles, then these moduli spaces are all topologically the same (namely $S^2=\PP^1$), and so counting BPS states is thus reduced to counting the number of these moduli spaces.\footnote{In order to avoid dealing with bound states we focus on cycles in primitive homology classes.} Finally, under the genericity assumption of \cite{filip:counting} which we detail below, given such a moduli space we can canonically realize our K3 surface as a special Lagrangian fibration by fibering the curve at a point in the moduli space (which will be singular at 24 points) over the base $S^2$.\footnote{Alternatively fibering the dual torus, or Jacobian, over the same base gives the Strominger-Yau-Zaslow construction of the mirror manifold \cite{strominger:mirrorT} which was studied for K3 surfaces in \cite{vafa:k3inst}.} Conversely, any special Lagrangian fibration of a K3 surface has torus fibers (plus 24 singular fibers) and a spherical base. So, we have rephrased our physics problem --
counting BPS states arising from M2-branes wrapping two-tori in the internal dimensions -- as one of counting special Lagrangian fibrations of a fixed K3 surface.

\bigskip
The organization of this note is as follows.  In \S2, we give a slightly more precise formulation of the counting problem.  In \S3, we 
solve the problem by a distinct method from \cite{filip:counting}.  In \S4, we use similar ideas to determine large-length asymptotics of closed billiard trajectories
on rectangular billiard tables, and briefly sketch the relation of this question to asymptotics of BPS state counts in 4d ${\cal N}=2$ gauge theories.

\section{More precise statement of the problem}

In order to be more explicit, we introduce the notion of a generic twistor family of marked K3 surfaces, $X$. `Marked' means that we have chosen an isomorphism $H^2(X,\ZZ)\to \Gamma^{3,19}$, where $\Gamma^{3,19}$ is the even unimodular lattice of signature $(3,19)$. Physically, this marking corresponds to a choice of U-duality frame. (Physicists may appreciate the discussion in \cite{Aspinwall:1996mn}.) A `twistor family' is obtained by taking a marked K3 surface, considering the positive-definite 3-plane $P\subset \Gamma^{3,19} \otimes \RR$ that contains the K\"ahler form and the real and imaginary parts of the holomorphic $(2,0)$-form, $\Omega$ (all of which we normalize to be unit vectors, so that they form an orthonormal basis for $P$), and rotating the K\"ahler form throughout the unit 2-sphere $S\subset P$. So, it is specified by a point $P$ in the Grassmannian 
\be
G_{3,19}=O(3,19)/O(3)\times O(19)
\ee
consisting of positive-definite 3-planes in $\RR^{3,19}$. All of the elements of the twistor family have the same Ricci-flat metric. Finally, `generic' means that all elements of the family have a N\'eron-Severi lattice, $H^2(X,\ZZ)\cap H^{1,1}(X)$, of rank (or `Picard number') $\rho$ at most one. Note that it is exactly this lattice that comprises the cohomology classes of holomorphic curves, so our genericity hypothesis is that for any choice of complex structure we have at most one primitive curve class. Moreover, a K\"ahler form, if one exists, must lie within this lattice, and so a K3 is algebraic if and only if this lattice is nontrivial. Note that the name `generic' is reasonable as such families correspond to the complement of a countable set of proper submanifolds of $G_{3,19}$ by lattice-theoretic considerations.

\bigskip
Now, more explicitly, the problem with which \cite{filip:counting, bergeron:slag} were concerned was to count the number of special Lagrangian fibrations in a generic twistor family of marked K3 surfaces whose non-singular fibers have volume at most $V_*$, in the limit of large $V_*$. Note that in a given fibration all of these fibers have the same volume as they are homologous cycles calibrated by $\Omega$.

\bigskip
In fact, we can easily work at the level of cohomology. That is, for any $Q\in H^2(X,\ZZ)$ with $Q^2\ge -2$ there is a K\"ahler form $\omega(Q)$ for which a representative of $Q$ is a special Lagrangian genus $Q^2/2+1$ cycle, and so in particular for any primitive null (or isotropic) $Q\in H^2(X,\ZZ)$ one obtains (under the genericity assumption) a special Lagrangian fibration. Indeed, this is a one-to-one correspondence \cite{filip:counting}. By using our freedom to choose the phase of $\Omega$, we can write the volume of a special Lagrangian representative of $Q$ as $Q\cdot\Omega(Q)$, where $\Omega(Q)$ is the holomorphic $(2,0)$-form corresponding to the K\"ahler form $\omega(Q)$.

\bigskip
Finally, we observe that the physical setup motivates the generalization to counting primitive null cohomology classes even when $P$ is not generic, although in this case the count is no longer one of special Lagrangian fibrations.

\bigskip
Having said all of this, we will now prove the following theorem at a physical level of rigor:
\begin{theorem}
Given a twistor family $X$ associated to a (possibly non-generic) 3-plane $P$ with unit sphere $S$, and given any smooth weight function $w: S \to \RR$, we have the following asymptotic formula, where the sum runs over all primitive null vectors $Q\in \Gamma^{3,19}$ with $Q\cdot \Omega(Q)\le V_*$:
\be \sum w(\Real\Omega(Q)) \sim \parens{\int_S d\Real\Omega\, w(\Real \Omega)} V_*^{20} \ . \label{eq:thm1}\ee
The implied constant multiplying the right hand side is independent of $w$.
\label{thm:1}
\end{theorem}
\noindent We also give a heuristic argument for the result of \cite{filip:counting} that the implied constant on the right side is independent of $P$, for generic $P$. The weight function is introduced in order to demonstrate that the points we are counting are equidistributed along $S$.

\bigskip
We prove this theorem by adapting methods of Douglas and collaborators used to count flux vacua and attractor black holes \cite{douglas:statistics,douglas:countingFlux,douglas:critical,douglas:distributions,douglas:nonSUSY,douglas:algGeom}.\footnote{As will be clear, this is a bit overkill. We use their methods both to stress the similarity of our counting problem with theirs and because it is neat to see their application in simple examples. For instance, in \S\ref{sec:pool} we will explain how to compute the area of a disk via an exponentially decaying integral over the whole complex plane! \label{ft:fun}} The restriction in the sum of \eqref{eq:thm1} to primitive vectors might seem troublesome, but since we are only interested in the large $V_*$ asymptotics we can essentially ignore it, since the probability for $n$ randomly chosen integers to have greatest common divisor 1 is $1/\zeta(n)$, which for $n=\dim H^2(X,\ZZ)=22$ is
\be \frac{1}{\zeta(22)} = \frac{13447856940643125}{155366 \pi ^{22}} \approx 1 \ .\ee
More precisely, for $n\ge 2$ integers selected uniformly at random in the range $[1,N]$, the probability for them to be relatively prime tends to $1/\zeta(n)$ as $N\to\infty$ \cite{nymann:relPrime}. Clearly, finite $N$ is correlated with finite $V_*$, and so the finite $N$ errors in this result will lead to finite $V_*$ errors. For this purpose, the more precise result $\frac{1}{\zeta(n)} + \Oo(1/N)$, valid for $n\ge 3$ \cite{nymann:relPrime}, will be useful.

\bigskip
Before continuing, we provide a simple proof that the count is finite. We decompose a null vector $Q\in \Gamma^{3,19}$ as $Q=Q_L + Q_R$, where $Q_L\in P$ and $Q_R \in P^\perp$. $Q^2=0$ implies $Q_L^2=|Q_R^2|$. The special Lagrangian condition (and our choice of phase for $\Omega(Q)$) forces $Q_L$ to be parallel to $\Real\Omega$. So, $Q\cdot\Omega = Q_L\cdot\Omega = \|Q_L\| = \|Q_R\| \le V_*$, and this restricts our attention to a finite set of vectors in $\Gamma^{3,19}$.

\bigskip
Similar reasoning allows one to show that
\be \sup_{t\in P, t^2=1} Q\cdot t = Q\cdot \Omega(Q) \ ,\ee
which relates our statement of Theorem \ref{thm:1} to that at the end of \S1.2 of \cite{filip:counting}.

\section{Proof}

We introduce the notation 
\be
\delta_\Omega = \frac{1}{2\pi} \delta\parens{\frac{Q_L\cdot\Real\Omega}{\|Q_L\|} - 1}
\ee
 for a delta function that forces $\Real\Omega$ to be parallel to $Q_L$, so that $Q$ has a special Lagrangian representative whose volume is $Q\cdot\Omega$.\footnote{This is normalized so that $\int_S d\Real\Omega \, \delta_\Omega f(\Real\Omega) = f\parens{\frac{Q_L}{\|Q_L\|}}.$} Similarly, 
\be
\delta_0=\frac{\Delta}{2 \|Q_L\|} \delta(Q^2)
\ee
imposes $Q^2=0$.\footnote{This is normalized so that $\int_{\RR^{0,19}} dQ_R\, \delta_0 = \Delta \|Q_L\|^{18} \int_{S^{18}} d\hat Q_R$\,, where $S^{18}$ is the unit 18-sphere parametrized by $\hat Q_R$, and $\int_{\RR^{3,0}} dQ_L\, \delta_0=\Delta \|Q_R\|^2 \int_{S} d\hat Q_L$.\label{ft:norm0}} Here, $\Delta$ is the width of our hyperboloid in $Q$-space, which is needed to appropriately enclose lattice points in the integral approximation below. Its most important feature is that as we rescale $Q\to Q/\alpha$, we will rescale $\Delta$ in the same way. 
Explicitly, $\Delta$ is determined by setting
\be
\sum_{\substack{Q \in \Gamma^{3,19}, ~Q^2=0,\\ \|Q_L\| \leq V_*}} ~\simeq~ \int_{\substack{{\mathbb R}^{3,19},\\ \|Q_L\| \leq V_*}} dQ ~\delta_0~, \label{eq:approx}
\ee
as $V_* \to \infty$.

\bigskip
In terms of these objects, we can estimate the count we are interested in (roughly by approximating a number of enclosed latticed points via a volume of the enclosing region) via
\begin{align}
\sum w(\Real\Omega) &\approx \frac{1}{\zeta(22)} \sum_{
\substack{Q\in \Gamma^{3,19},\\ Q^2=0}} \int_S d\Real\Omega\, \delta_\Omega \, w(\Real\Omega) \, \theta(V_*-Q\cdot\Omega) \label{eq:approx1} \\
&\approx \frac{1}{\zeta(22)} \int_{\RR^{3,19}} dQ\, \delta_0 \int_{S} d\Real\Omega\, \delta_\Omega \, w(\Real\Omega) \, \theta(V_* - Q\cdot\Omega)
 \label{eq:approx2} \ .
 \end{align}
Next, we re-write the theta function in terms of a contour integral, so that \eqref{eq:approx2} becomes
\be
\lim_{\epsilon\to 0^+} \frac{1}{2\pi i} \frac{1}{\zeta(22)} \int_{\epsilon-i\infty}^{\epsilon+i\infty} \frac{d\alpha}{\alpha}\,e^{\alpha V_*} \int_{\RR^{3,19}} dQ\, \delta_0 \int_{S} d\Real\Omega\, \delta_\Omega \, w(\Real\Omega) \, e^{-\alpha \, Q\cdot\Omega} 
 \ .
\ee
Now, we rescale $Q\to Q/\alpha$.\footnote{This is justified in more detail in \cite{mz:nonSUSYattr}.} The $\alpha$ integral then gives
\begin{align*}
\lim_{\epsilon\to 0^+} \frac{1}{2\pi i} \int_{\epsilon-i\infty}^{\epsilon+i\infty} \frac{d\alpha}{\alpha^{21}} e^{\alpha V_*} &= \frac{V_*^{20}}{20!} \ .
\end{align*}
So, we get
\be \frac{V_*^{20}}{20!} \frac{1}{\zeta(22)} \int_{\RR^{3,19}} dQ\, \delta_0 \int_{S} d\Real\Omega\, \delta_\Omega \, w(\Real\Omega) \, e^{-Q\cdot\Omega} \ .\ee

\bigskip
Performing the radial $Q_L$ integral using $\delta_0$, the angular integral over $\hat Q_L = Q_L/\|Q_L\|$ using $\delta_\Omega$, and then the $Q_R$ integral gives
\begin{align}
\int_{\RR^{3,19}} dQ\, \delta_0\, \delta_\Omega \, e^{-Q\cdot\Omega} &= \Delta\int_S d\hat Q_L \, \int_{\RR^{0,19}} dQ_R\, \|Q_R\|^2 \delta_\Omega e^{-(\hat Q_L\cdot \Real\Omega)\,\|Q_R\|} \\
&= \Delta \int_{\RR^{0,19}} dQ_R\, \|Q_R\|^2e^{-\|Q_R\|} \\
&= \Delta \cdot \frac{2\pi^{19/2}}{\Gamma(19/2)} \int_0^\infty dr\, r^{20} e^{-r} \\
&= \Delta \cdot 72296379187200 \, \pi^9 \ .
\end{align}
So, the implied constant on the right hand side of \eqref{eq:thm1} is
\be \Delta\cdot \frac{199808985600}{77683 \pi ^{13}} \ .\ee
This result should agree with the ratio of volumes of homogeneous
spaces appearing in the constant of \cite{filip:counting}. In particular, $\Delta$ should be independent of $P$. While the analogous statement manifestly fails for the even unimodular lattice $\Gamma^{1,1}$, we expect it to hold away from low dimensions, as the integrals over the left- and right-moving unit spheres should wash away dependence on $P$.

\bigskip
Clearly, the dominant source of error in the approximation \eqref{eq:approx2} in the large $V_*$ limit comes from the boundary, $\|Q_L\|=\|Q_R\|=V_*$. We have already discussed the errors in \eqref{eq:approx1}. It would be interesting to compare these errors to those discussed in \cite{bergeron:slag}.

\section{Billiards} \label{sec:pool}

We now demonstrate that identical techniques can be applied to billiards problems of interest in the study of flat surfaces, or translation surfaces, which are formally analogous to the counting problem treated above \cite{filip:counting}. We refer the reader to \cite{zorich:flat} for background. Here, we will focus on the simple problem of counting closed geodesics on a genus one surface whose area we normalize to unity.

\bigskip
We regard the torus as arising from the identification of opposite edges of a parallelogram. We then tile the complex plane with translates of this parallelogram and observe that the corners thereof form a unimodular lattice, $\Gamma\in \CC$. We choose the origin to coincide with one such corner. Now, our problem is reduced to counting geodesics in $\CC$ which begin at the origin and end at a primitive vector in $\Gamma$.  
We note that when the parallelogram is rectangular, this is equivalent to counting closed geodesics on a billiards table. 
See the illustration in Figure 1.

\begin{figure}[h!]
  \caption{A geodesic on a billiard table; its lift to a tiling of the plane; and the same trajectory viewed as a geodesic on a surface of genus 1.}
  \label{fig:whatthefuck}
  \centering
    \includegraphics[width=0.3\textwidth]{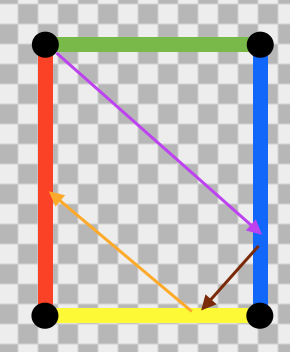}
\includegraphics[width=0.3\textwidth]{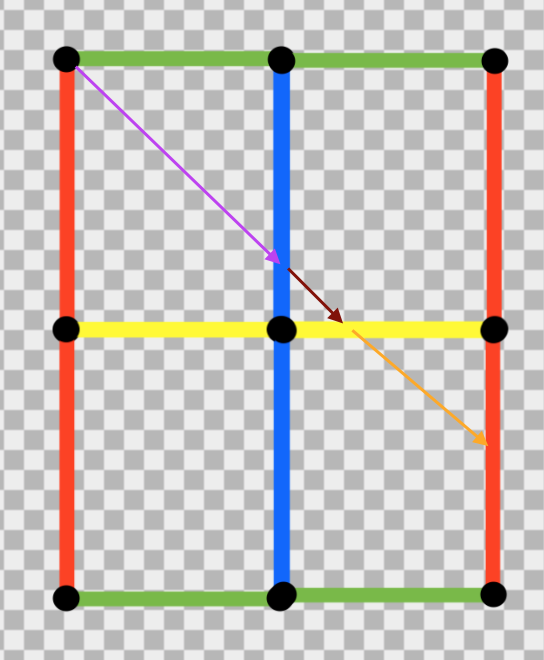}
\includegraphics[width=0.3\textwidth]{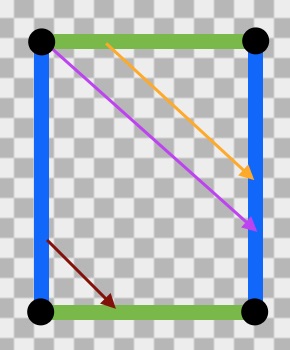}
\end{figure}

\bigskip
We denote both the angle of the geodesic relative to the horizontal and the unit vector in this direction by $\theta$. Let $\delta_\theta=\delta\parens{\theta-\arg Q}$ be a delta function that forces $\theta$ to be parallel to $Q$ and let $S$ denote the unit circle. Then, the count is
\begin{align}
\sum w(\theta) &\approx \frac{1}{\zeta(2)} \sum_{Q\in \Gamma} \int_S d\theta\, \delta_\theta \, w(\theta) \theta(L_* - Q\cdot \theta) \\
&\approx \frac{1}{\zeta(2)} \int_{\CC} dQ\, \int_S d\theta\, \delta_\theta \, w(\theta)\theta(L_* - Q\cdot\theta) \ ,
\end{align}
where the $\sum$ runs over geodesics ending at primitive lattice points, and $w(\theta)$ is an arbitrary smooth function.

\bigskip
We now do the same tricks as before to deal with the $\theta$ function, so that we are left with
\begin{align}
\sum w(\theta) &\approx \frac{L_*^2}{2!} \frac{1}{\zeta(2)} \int_{\CC}dQ\, \int_S d\theta\, \delta_\theta \, w(\theta) e^{-Q\cdot\theta} \\
&= \frac{3L_*^2}{\pi^2} \int_{\RR^+}dr\, \int_S d\theta\, w(\theta) r e^{-r} \\
&= \frac{3 L_*^2}{\pi^2} \int_S d\theta\, w(\theta) \ .
\end{align}
In particular, setting $w=1$ and performing the integral yields $\frac{6}{\pi^2}\cdot \pi L_*^2$, which shows that the Siegel-Veech constant for $T^2$ is $\frac{6}{\pi^2}$. We note that this quadratic asymptotic behavior applies to a very general class of geodesic counting problems \cite{veech:billiards,masur:geodesics1,masur:geodesics2,vorobets:geodesics1,eskin:billiards1}.

\bigskip
We can also relate this sort of counting problem to physics.\footnote{`Sort of' refers to the fact that the relevant surfaces in the physics problem are generally not as nice as those considered above. In particular, they have infinite area, since the Seiberg-Witten differential is only meromorphic, as opposed to holomorphic. Nevertheless, counts of geodesics on such surfaces have been investigated recently \cite{tahar:poles}. The relationship between geodesics and BPS states is discussed in more detail, and exploited in order to determine asymptotics of BPS spectra, in \cite{mz:asymp}; in particular, examples with holomorphic differentials are discussed. E.g., counting geodesics on an elliptic curve is related to the BPS spectrum of the $\N=4$ $SU(2)$ theory.} Specifically, consider the Seiberg-Witten description of the low-energy limit of an $SU(2)$ gauge theory, possibly with matter in the fundamental representation \cite{sw:theory1,sw}. This involves an elliptic fibration over the $\PP^1$ Coulomb branch of the gauge theory. We view a fiber, called the Seiberg-Witten curve, $\Sigma$, as being cut out of $\PP^2$, with coordinates $x,z$, by a single equation. This embedding is quite useful because a number of properties of the low-energy physics are illuminated by regarding the Seiberg-Witten curve as a branched double cover of the $z$-plane. (This can be motivated by geometrically engineering the theory in string theory, where the $z$-plane arises as the base of an ALE-fibration \cite{kachru:geomEng,vafa:geodesics,brand:geodesics}.) In particular, $\Sigma$ is endowed with a canonical flat surface with poles structure \cite{boissy:poles,tahar:poles} by a meromorphic one-form, $\lambda$, known as the Seiberg-Witten differential. (Note that this induces a rather different metric, $g_{z\bar z}=4\lambda_z\bar\lambda_{\bar z}$, on $\Sigma$ from the one obtained from its embedding in $\PP^2$.) Stable BPS hypermultiplets then correspond to primitive geodesics connecting distinct branch points associated to zeroes of $\lambda$, while vector multiplets correspond to primitive closed geodesics \cite{vafa:geodesics,brand:geodesics}.\footnote{This is clearly related to the discussions in  \cite{bs:quadStab,kontsevich:flatStable}; see for instance the opening paragraph in \S1.6 of \cite{bs:quadStab}.}

\bigskip
In fact, this same 4d $\N=2$ field theory may be engineered in another way in string theory, and this leads us to another math problem which is dual to the one just discussed. Specifically, these field theories describe the worldvolume of a D3-brane probing D7-branes and an O7 orientifold plane \cite{s:fBranes}. Quantum corrections to this picture are described by F-theory on an elliptically fibered K3 surface \cite{sen:FOrientifolds}, where the Coulomb branch is an infinitesimal neighborhood in the base of the fibration. BPS states now correspond to certain geodesics, and webs thereof, on this base \cite{bergman:FWebs,sethi:FWebs}; these geodesics trace out the positions of strings.

\bigskip
We thus see that two different counting problems involving geodesics on different surfaces have the same answers. The connection between them is provided by compactifying the F-theory setup on a circle transverse to the D3-brane, T-dualizing, and lifting to M-theory. The compactification manifold is our original K3 surface, but now the fibers are also part of spacetime, and their K\"ahler class is controlled by the radius of the extra circle we introduced in the F-theory frame. The D3-brane becomes an M5-brane wrapping these fibers \cite{w:MGaugeSol}, while BPS states are now M2-branes ending on the M5-brane with one leg along the base and one leg along the fibers \cite{yi:surfaces, mikhailov:surface}. It is now clear that each of the earlier counting problems was just looking at one of the two legs of the M2-brane.

\bigskip
\section*{Acknowledgments}

The research of S.K. was supported in part by a Simons Investigator Award and the National Science Foundation under grant number PHY-1720397.  The research of A.T. was supported by the National Science Foundation under NSF MSPRF grant number 
1705008.

\bibliography{Refs}

\end{document}